\documentclass{article}

\usepackage{arxiv}

\usepackage[utf8]{inputenc} 
\usepackage[T1]{fontenc}    
\usepackage{hyperref}       
\usepackage{url}            
\usepackage{booktabs}       
\usepackage{amsfonts}       
\usepackage{nicefrac}       
\usepackage{microtype}      

\usepackage{graphicx}
\usepackage{natbib}
\usepackage{doi}

\usepackage[utf8]{inputenc}
\usepackage{amsmath}

\date{} 					
\title{Interpretable Neural Networks to Predict Momentum Fluxes of Orographic Gravity Waves}

\author{ 
    \href{https://orcid.org/0009-0004-1265-3209}{\includegraphics[scale=0.06]{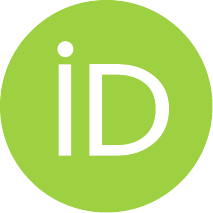}\hspace{1mm}Elias Haslauer\textsuperscript{1,}}\thanks{\texttt{elias.haslauer@dlr.de}} \\
    \And
    \href{https://orcid.org/0000-0001-6565-5890}{\includegraphics[scale=0.06]{orcid.pdf}\hspace{1mm}Mierk Schwabe\textsuperscript{1}} \\
    \And
    \href{https://orcid.org/0000-0003-0936-0216}{\includegraphics[scale=0.06]{orcid.pdf}\hspace{1mm}Andreas Dörnbrack\textsuperscript{1}} \\
    \And
    \href{https://orcid.org/0000-0002-6010-6638}{\includegraphics[scale=0.06]{orcid.pdf}\hspace{1mm}Edwin P. Gerber\textsuperscript{2}} \\
    \And
    \href{https://orcid.org/0000-0003-1508-5900}{\includegraphics[scale=0.06]{orcid.pdf}\hspace{1mm}Markus Rapp\textsuperscript{1,3}} \\
    \And
    \href{https://orcid.org/0000-0002-7256-5073}{\includegraphics[scale=0.06]{orcid.pdf}\hspace{1mm}Nedjeljka \v{Z}agar\textsuperscript{4}} \\
    \And
    \href{https://orcid.org/0000-0002-6887-4885}{\includegraphics[scale=0.06]{orcid.pdf}\hspace{1mm}Veronika Eyring\textsuperscript{1,5}} 
}

\affiliations{
    \textsuperscript{1}Deutsches Zentrum für Luft- und Raumfahrt e.V., Institut für Physik der Atmosphäre, Germany\\
    \textsuperscript{2}Courant Institute, New York University, New York, USA\\
    \textsuperscript{3}Meteorologisches Institut, Ludwig-Maximilians-Universität München, Germany\\
    \textsuperscript{4}Meteorological Institute, University of Hamburg, Germany\\
    \textsuperscript{5}Institute of Environmental Physics (IUP), University of Bremen, Germany
}

\date{}

\begin{document}
\maketitle

\begin{abstract}
State-of-the-art Earth system models (ESMs) cannot explicitly resolve many small-scale atmospheric processes such as atmospheric gravity waves, and thus must represent, or parameterise, their effects on the resolved state. Machine learning (ML) has the potential to improve these parameterisations. In our study, we train neural networks (NNs) on ERA5 reanalysis data to predict momentum fluxes of orographic gravity waves as a function of the state variables at the resolution of a coarse ESM. Employing a full year of data, we extract inertia-gravity waves using the software MODES, which applies linear theory for wave filtering, and train ML models on data coarse-grained to the ESM's target resolution. We consider four different cases: the full spectrum of inertia-gravity waves resolved in ERA5, or just the part of the spectrum that is subgrid-scale in the target ESM, both over all land or just over mountainous terrain. Our NNs successfully predict momentum fluxes, with a global coefficient of determination ($R^2$) ranging from 0.72 to 0.56, depending on the case, when evaluated offline with data from another year. An analysis of our models using SHAP values, an explainable AI technique, suggests that the networks learned physically meaningful relationships. In addition, we give a comparison with the physics-based parameterisation scheme by Lott and Miller. This work forms the basis for the development of operational ML-based parameterisations to improve the representation of gravity waves and their effects in climate models.
\end{abstract}


\section{Introduction}

Earth system models (ESMs) are key to a better understanding of the Earth system, projecting future climate change, and deriving suitable strategies for mitigation and adaptation in response to global warming. While models have significantly improved over the last decades regarding the simulation of the mean climate and its variability for many large-scale indicators of climate change – mainly due to 
increasing model resolutions and a better representation of physical processes – they still face systematic errors and large uncertainties \citep{Beverly_2024, Eyring_2024, Vautard_2023, Vicente-Serrano_2022}. A main reason for these uncertainties is the difficulty to represent small-scale processes like clouds, convection, and microphysics, processes that cannot be captured explicitly by state-of-the-art models, which typically run at grid resolutions of around 50 to 100 km. The statistical effects of these subgrid processes are represented by so-called parameterisations, which are traditionally based on physical process understanding and empirical relationships, but often come along with severe simplifications necessary to make them computationally efficient \citep{Achatz_2024, Eyring_2024, Kim_2003, Stensrud_2007}.
  
Gravity waves are atmospheric disturbances in which buoyancy acts on air parcels displaced from hydrostatic equilibrium. They are called inertia-gravity waves if the waves are of large scale, such that the Coriolis force has substantial additional influence. These waves constitute an important class of subgrid-scale processes \citep{Fritts_2003}. Gravity waves have traditionally been differentiated depending on their sources: orographic gravity waves arising from wind flowing over mountains, vs. non-orographic gravity waves, induced by, e.g., convection, jets, and fronts. As gravity waves propagate through the atmosphere, they transport momentum, and influence the atmospheric circulation when they finally break and deposit their momentum. Since gravity waves occur at any scale, with horizontal wavelengths ranging from hundreds of metres to several thousands of kilometres, a significant part of their spectrum cannot be captured at the resolution of today’s climate models. Even though the evolution of computational capabilities allows the integration of global storm resolving models with resolutions on the order of kilometres, gravity wave parameterisations are still needed since even these highly resolved models don't capture the smallest scales of gravity waves, and coarse resolution models are needed, e.g., to run large model ensembles \citep{Achatz_2024, Polichtchouk_2023}.

Initially, gravity wave parameterisations were introduced to correct for a westerly bias in models \citep{Palmer_1986}, improving their numerical stability and prediction skill. However, even today's operational gravity wave parameterisations (e.g., \cite{Hines_1997, Lott_1997, Lott_1999}) operate with oversimplifying assumptions, the severest being neglecting horizontal propagation \citep{Achatz_2024, Alexander_2010, Eichinger_2023, Stephan_2019}. The consequences of these shortfalls include, e.g., an unrealistic or absent quasi-biennial oscillation in many models (QBO, \cite{Richter_2022}), missing gravity wave drag over the Southern Ocean \citep{McLandress_2012}, and deficiencies in the simulation of the wintertime polar vortex and sudden stratospheric warmings (SSWs, \cite{McLandress_2013}). These issues can partly be addressed by tuning. Also, it is an open question whether the models will generalise in a changing climate \citep{Achatz_2024}. Hence, improvements in accuracy and physical adequacy are highly desirable.

In the last years, much research has been conducted to replace conventional parameterisation schemes in climate models with machine learning (ML) approaches (e.g., \cite{Gentine_2018, Grundner_2022, Heuer_2024, Rasp_2018, Sarauer_2025, Yuval_2020, Yuval_2023}, for an overview see \cite{Burgh_2023}). ML techniques enable the description of complex non-linear relationships based on data and have revolutionised many fields of science in the last decade. Neural networks (NNs) are trained on large amounts of “known” data, allowing predictions also for unseen data. For a general review of ML, deep learning techniques, and terminology see \cite{Alzubaidi_2021}.

In the context of ML-based parameterisations, there are two main approaches: first, emulations, which mimic conventional schemes with the goal of reducing the computational requirements. Second, training ML models with high-resolution simulations or observations that resolve the process of interest. In regard to the parameterisation of gravity waves, \cite{Chantry_2021}, \cite{Connelly_2024}, \cite{Espinosa_2022},
\cite{Hardiman_2023}, and \cite{Sun_2024} followed the first strategy, emulating existing schemes. While these ML-based parameterisations might imitate their reference models adequately and improve computational performance, they inherit the limitations of traditional gravity wave parameterisations discussed above.

Related to the second approach, NNs have been applied using reanalysis data to predict gravity wave momentum fluxes (GWMFs) locally over Japan \citep{Matsuoka_2020}, as well as non-orographic gravity waves over sea \citep{Amiramjadi_2023}. \citet{Dong_2023} train an ML model on high-resolution data to speed up a physics-based model. \citet{Gupta_2024a} aim for a global approach, also including the representation of non-local gravity wave effects by taking into account data of neighbouring grid cells, as well of the whole globe. This is further pursued in \citet{Gupta_2025}, where the authors predict GWMFs by fine-tuning a foundation model pretrained for weather and climate applications.

In this work, we train NNs on GWMFs of atmospheric reanalysis data, and successfully predict fluxes associated with mesoscale inertia-gravity waves based on the coarse state variables. As a first step, we focus primarily on orographic gravity waves, since their subgrid sources can be identified easily. To this end, the ERA5 global reanalysis dataset \citep{Hersbach_2020} is used as high-resolution training data. Since ERA5's resolution captures only part of the gravity wave spectrum, this is only a first step towards using higher resolution training data in the future.

Gravity waves are extracted using the spherical linear theory for the decomposition of three-dimensional circulation \citep{Kasahara_1981} with the software MODES \citep{Zagar_2015}. We investigate both the case of the full spectrum of gravity waves (IG, i.e., all inertia-gravity waves), as well as a limited part of the spectrum which is ``subgrid" (SG), i.e., unresolved at a resolution as it would be used in a climate model. After the computation of the respective momentum fluxes, the data are coarse-grained to this lower resolution, in which we train a modified U-Net architecture \citep{Ronneberger_2015} on atmospheric columns. The results are physically interpretable when analysing SHAP values (SHapley Addiditve exPlanations, \cite{Lundberg_2017}), an explainable AI (XAI) technique, and yield encouraging results when comparing to the conventional gravity wave parameterisation scheme by Lott and Miller \citep{Lott_1997, Lott_1999}, which is used operationally in many weather and climate models.

The remainder of this paper is structured as follows: Section 2 describes the dataset used in this study, our method of extracting gravity waves and calculating corresponding momentum fluxes, the architecture of the NNs, and the different experiments of this study. We also give a short overview of SHAP values. In Section 3, we present offline (i.e., the parameterisation is not coupled to a climate model) results of the NNs, the analysis of SHAP values, and a comparison of our networks with the conventional Lott and Miller scheme. We briefly discuss our results in Section 4. Section 5 gives an outlook and sketches the road ahead, in particular extending the approach to non-orographic gravity waves and adapting the NNs for the use in the ICON XPP model \citep{Mueller_2025}.

\section{Data and Methods}

\subsection{Dataset and General Approach}

This study uses the ERA5 global reanalysis dataset provided by the European Centre for Medium-Range Weather Forecasts (ECMWF, \cite{Hersbach_2020}). ERA5 is produced with the ECMWF Integrated Forecasting System (IFS) Cy41r2 on 137 hybrid $\sigma$-pressure levels with a model top at 0.01\,hPa and $\sim31\,\mathrm{km}$ horizontal resolution (TL639). We use the version provided on 37 vertical pressure levels up to 1\,hPa, with a horizontal resolution of $0.25^{\circ} \times  0.25^{\circ}$ (i.e., a grid spacing of $\sim28\,\mathrm{km}$ at the equator), and the time resolution of 1 hour. The full year 2024 is used for training of the NNs, while days 1, 11, and 21 of each month of the year 2022 serve as a test set. Such an amount of training data and its distribution over one year is needed in light of the seasonal variability of numerous phenomena associated with gravity waves, including sudden stratospheric warmings, and the changes in atmospheric dynamics over the year. We found that the NNs showed a substantial decline in their ability to predict momentum fluxes when trained with less data.

We are interested in the prediction of GWMFs of both the full spectrum of gravity waves, as well as of subgrid-scale GWMFs, as a function of the coarse state variables. Therefore, our strategy is as follows: The original ERA5 data, given on a $720 \times 1440$ latitude-longitude grid, are considered as high-resolution ground truth for the purpose of this study. Approximately eight grid points are needed to resolve a gravity wave adequately, which determines the effective resolution of the grid \citep{Sun_2023}. In our case, waves with horizontal wavelengths of $\sim200\,\mathrm{km}$ and more are fully resolved \citep{Gupta_2024b}. This means that we are dealing with mesoscale inertia-gravity waves and are missing a substantial fraction of the gravity wave spectrum in this initial work.

We train the NNs at the coarser target resolution of $64 \times 128$ grid points ($\sim300\,\mathrm{km}$ at the equator). At that resolution, only gravity waves of scale $\sim2000\,\mathrm{km}$ and larger are fully resolved. Thus, it is possible to regard the ``small-scale" part of the gravity wave spectrum (wavelengths between $\sim200\,\mathrm{km}$ and $\sim2000\,\mathrm{km}$) which is resolved in the high-resolution data, but not at the coarse resolution, as the subgrid part. This part needs to be parameterised in a climate model running at the coarse resolution. Such an approach will allow replacing the conventional gravity wave parameterisation in a coarse model in future work. Note that all statements made here about resolution depend on the latitude when using a lat-lon grid, and the values given above are estimated at the equator. Regarding the vertical resolution, we keep all 37 pressure levels.

\subsection{Extraction of Gravity Waves Using Normal Mode Functions}

A crucial decision is how to identify and separate the gravity waves from the global dynamical fields, and the filtering for specific wavelengths. \citet{Sun_2023} give insight into different methods for the extraction of gravity waves. Following a different approach, we use the software MODES \citep{Zagar_2015} which is based on the theory of normal-mode function expansion \citep{Hough_1898, Longuet-Higgins_1968, Kasahara_1981}.

Normal-mode functions constitute a spectral basis in which the global wind and mass fields can be expressed simultaneously, and permit the decomposition into two basic types of motion, inertia-gravity waves and Rossby waves, with different horizontal length scales and vertical structures. The calculation of this basis starts from a linearised system of primitive equations, which is used to obtain differential equations describing the vertical and horizontal structure of the atmosphere. The basis functions are eigensolutions of these differential equations. For a detailed mathematical derivation, we refer to \cite{Kasahara_1981} and \cite{Zagar_2015}. 

Once the horizontal and vertical basis functions are obtained for the given background global stability and vertical grid, the global circulation can be projected onto it. The expansion completeness supports filtering of the desired wave type (here just the gravity waves) and wavelengths to physical space to reconstruct the velocity and temperature perturbations of waves of interest. The filtering does not involve wave frequencies. This process is conducted at every time step independently using the precomputed basis functions. MODES gives a precise linear separation of inertia-gravity waves from the dynamical background, since the dispersion relations of the different wave types are inherent in the underlying theory of normal-mode functions. Although not of interest here, MODES gives both wind and temperature perturbations of gravity waves, since it is a multitvariate decomposition. For gravity wave wind and temperature filtering see for example \cite{Zagar_2017}.

We apply MODES at the full ERA5 resolution. To obtain the full spectrum of inertia-gravity waves (IG), all wavelengths of the inertia-gravity wave type are projected back to physical space; for the subgrid-scale part (SG) at our selected coarse resolution, we filter for all inertia-gravity waves with zonal wavenumber (i.e., the number of wavelengths fitting into a circle of latitude) greater than 16. This corresponds to wavelengths smaller than $\sim 2500\,\mathrm{km}$ at the equator, which fits well with the typically used cut-off scale in many gravity wave studies (e.g. \citet{Polichtchouk_2023}). This wavelength decreases as moving towards the poles, e.g., it is $\sim1200\,\mathrm{km}$ at 60° latitude. We consider these waves to be  unresolved by the coarse grid. The full set of parameters used for running MODES can be found in Appendix \ref{MODES_settings}.

The applied version of MODES operates on terrain-following $\sigma$-levels. We define $\sigma$-levels using our given 37 pressure levels, and interpolate the input ERA5 data from pressure to $\sigma$-levels. After applying the backward projection to filter gravity waves, results are interpolated from $\sigma$-levels back to pressure levels. The vertical interpolations are carried out using Climate Data Operators (CDO, \cite{Schulzweida_2023}) with the function \emph{intlevelx3d}. This is different from a typical MODES setup which operates on the hybrid $\sigma$-pressure model levels of the ECMWF IFS model and interpolates them to $\sigma$-levels. The additional interpolation steps involved in our case are not considered detrimental to the study focusing on the middle atmosphere. 
The new version of MODES operating in the pressure vertical coordinate \citep{Zagar_2023} will facilitate future work.

\subsection{Calculation and Evaluation of Gravity Wave Momentum Fluxes}\label{CalcFlux}

MODES yields global fields of horizontal wind perturbations $(u',v')$ corresponding to the inertia-gravity waves and (in the subgrid-scale case) the filtered, unresolved wavelengths. Since vertical velocities $\omega$ in ERA5 are given in $Pa/s$, we calculate zonal and meridional momentum fluxes using the formulas
\begin{align*}
    MF_x &= - g^{-1} \overline{u' \omega} \\
    MF_y &= - g^{-1} \overline{v' \omega}.
\end{align*}
Here, $g=9.81\, \mathrm{m/s}^2$ is the gravitational acceleration of Earth and $\overline{\ ^{ }\cdot^{ }\ }$ denotes averaging, which we implicitly perform by coarse-graining the data horizontally to a $64 \times 128$ grid with first order conservative remapping, using the CDO function \emph{remapcon}. The version of MODES used for this study does not provide the pressure velocity perturbations $\omega'$. Therefore, we use $\omega$ directly from ERA5, assuming that the true average is approximately zero (that is, $\omega' \gg \overline{\omega}$), and that in the areas occupied by gravity waves, $\omega$ stems mainly from small-scale motions associated with the waves and not from the large-scale balanced flow. This simplification follows \cite{Prochazkova_2025}. 

\begin{figure}[t!]%
\includegraphics[width=1.0\textwidth]{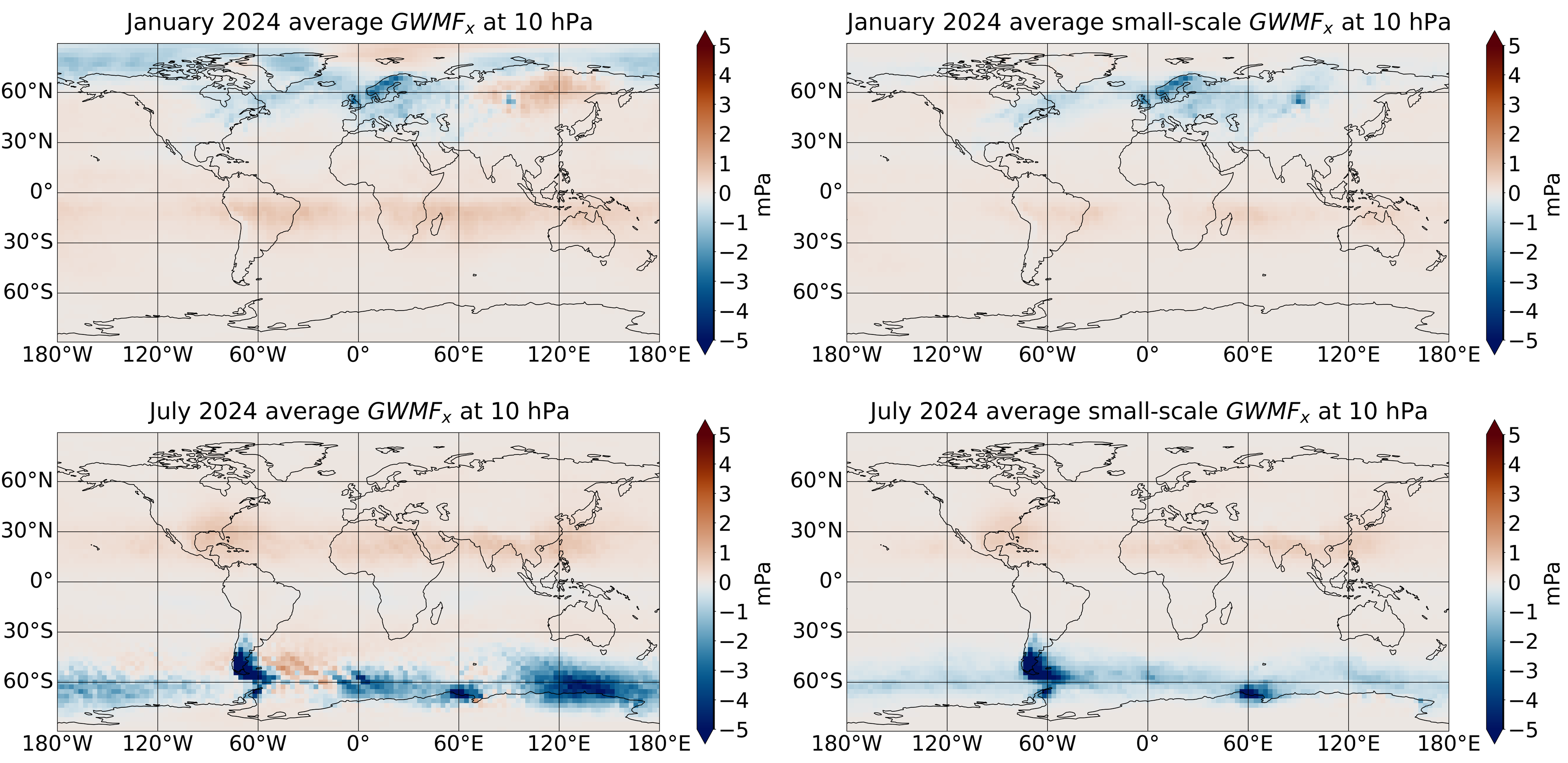}
{\caption{Monthly averages of zonal GWMFs in ERA5 for the full spectrum of gravity waves (left) and the small-scale part (right) at 10\,hPa, for January (top) and July (bottom) 2024.}
\label{fig_AvgFluxesIG}}
\end{figure}

Monthly averages of zonal momentum fluxes for January and July at 10\,hPa are shown in Figure \ref{fig_AvgFluxesIG} for the full spectrum and the subgrid-scale part, respectively. Monthly averages of the months January, April, July, and October 2024 can be found in the Appendix (Figures \ref{MF_IG_seasons} and \ref{MF_SG_seasons}). For further discussion of the dataset itself, additional figures and videos, as well as a detailed evaluation of the GWMFs in ERA5 calculated for this study, please see [anonymised, paper in preparation].

An evaluation of GWMFs with observational data is difficult. Since we consider global data, only estimations of fluxes based on temperature measurements with satellites provide the necessary spatial coverage. Valuable references are \citet{Ern_2011}, \citet{Geller_2013}, and \citet{Hindley_2020}, considering data collected by the HIRDLS, SABER, and AIRS satellites. However, quantitative comparison remains delicate, since data is from different years, and satellite measurements cover heights of $\sim30\mathrm{km}$ and above only. Here, we only give a very short summary and refer to [anonymised, paper in preparation] for more details.

Considering the full spectrum of inertia-gravity waves, the fluxes found with MODES in ERA5 on 1\,hPa and 10\,hPa are roughly of the same order of magnitude as those obtained from HIRDLS and SABER satellite measurements \citep{Ern_2011} for January and July, but are weaker on average. We attribute this to missing smaller scale waves in ERA5. This finding is consistent with prior analyses by, e.g., \citet{Lear_2024} and \citet{Yoshida_2024}. There is, however, a good qualitative agreement. For example, we observe relatively strong negative (in this case, indicating up- and eastward propagating) GWMFs between 50°S and 75°S in the months June, July, and August, which stem primarily from the wave activity associated with flow across the Andes mountains. Also, there is good qualitative agreement with the analysis of GWMFs conducted by \citet{Gupta_2024}.

Regarding the small-scale part of the spectrum, we found no appropriate data available to compare to. As expected, both mean and maxima of momentum fluxes associated with small-scale gravity waves are clearly smaller in magnitude than those of the full spectrum. However, the general patterns of the fluxes are similar to those of the full spectrum.

Taking a closer look at  Figure~\ref{fig_AvgFluxesIG}, the large-scale structure of fluxes especially at 10~hPa in the Southern Hemisphere deserves some attention, related to the linear decomposition and the use of the total $\omega$ field. Linear wave decomposition by MODES separates 3D circulation between geostrophically balanced Rossby waves on the sphere and the remaining signal projecting on IG modes. Linear, spherical Rossby waves have a small divergence associated with the beta term, proportional to $v_g\, \beta/f$, where $v_g$ is the meridional geostrophic wind, $f$ is the Coriolis parameter and $\beta$ is its meridional gradient. The isallobaric motions, i.e. most of ageostrophic dynamics related to the baroclinic Rossby wave dynamics in midlatitudes, projects on the IG modes.  This means that large-scale midlatitude IG modes are a mixture of ageostrophic circulation and inertia-gravity (or gravity) waves. The zonal wavenumber 3 structure of the flux at 10~hP in midlatitudes thus reflects ageostrohic dynamics associated with vertically-propagating Rossby waves in the winter hemisphere \citep[see discussion in][]{Zagar_2023}. After filtering out large scales, this part of the signal is greatly weakened.

\subsection{Architecture of Neural Networks}\label{NN_architecture}

As a first step towards developing physically consistent parameterisations for both orographic and non-orographic gravity waves, this study focuses primarily on orographic gravity waves. More precisely, we consider GWMFs only over land and use input features that describe the unresolved topography. NNs are trained on full atmospheric columns to incorporate the vertical propagation of gravity waves. The target variables of the networks are vectors of zonal and meridional momentum fluxes on 37 levels $\{\mathrm{MF}_x, \mathrm{MF}_y\}$, summing up to 74 variables in total. As feature variables, we use columns of three-dimensional wind and temperature $\{ u, v, \mathbf \omega, T \}$, and five scalar variables describing the orography within a grid cell, namely mean geopotential at the ground level $z$, standard deviation $\mu$, anisotropy $\gamma$, angle $\sigma$ (i.e., orientation of the terrain relative to an eastward axis), and slope $\theta$ of subgrid-scale orography. This amounts to 153 feature variables in total, all of which are obtained by coarse-graining ERA5 data.

Since we are interested in orographic gravity waves and do not consider horizontal propagation in the current study, we only use columns over land, i.e., all columns where the land-sea-mask of ERA5 is non-zero. Further, we exclude all columns north of 70°N and south of 70°S in the training data, to omit over-representation of these areas, where grid points lie increasingly dense. Starting from $64 \times 128 =$ 8,192 columns per time step, this reduction leaves us with 2,794 columns per time step, which we use as samples for training the NNs.

Harnessing the full year 2024 for training, this gives us 24,542,496 training samples. As gravity waves are strongly related to specific atmospheric conditions (such as strong winds flowing over mountains) which often endure for several hours or even days, picking training and test data randomly from the same year seems unsuitable. Consecutive time steps are not independent in the setup with a time resolution of 1 hour. Instead, to get a reliable estimate, we take the days 1, 11, and 21 of each month in the year 2022 as test set. This yields 2,414,016 test samples. Due to the high computational demands of data processing in this case, we did not prepare an independent validation set.

All variables are normalised level-wise by subtracting the mean and dividing by the standard deviation. While normalisation is a common step in many ML applications, we found it crucial for this task, since the magnitudes of momentum fluxes depend on the model level due to the decrease of density $\rho$ with height. Without normalisation, upper levels would matter less. To avoid this, we correct by normalising every level separately.

\begin{figure}[t!]%
\includegraphics[width=1.0\textwidth]{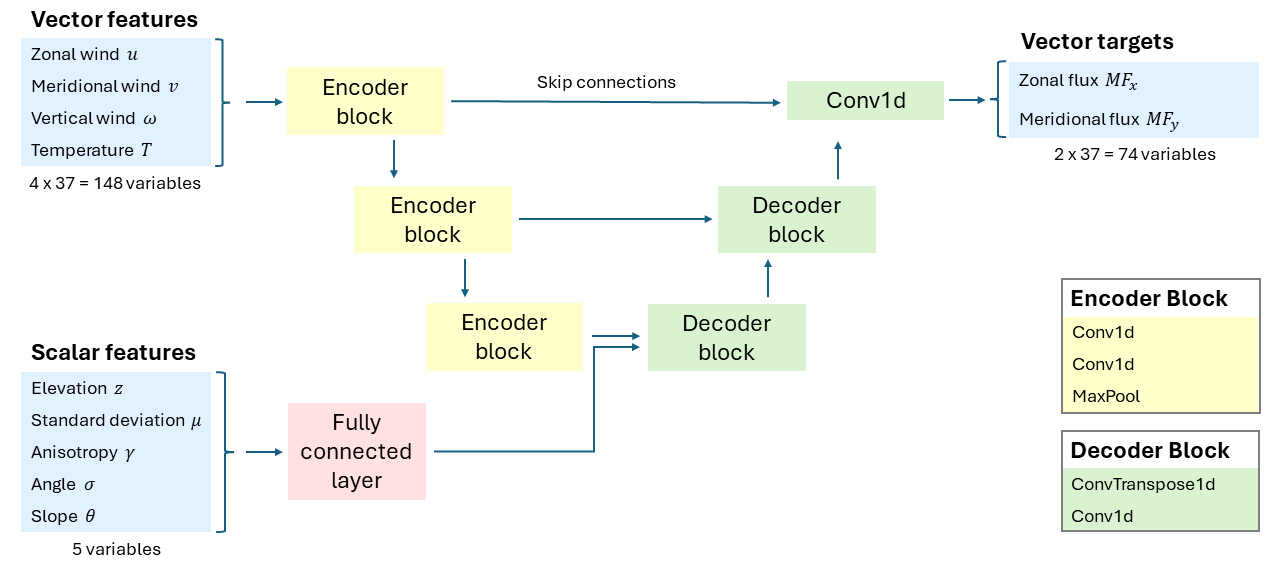}
{\caption{Modified U-Net architecture used in this study. First, vector features are passed through three encoder blocks. In the bottleneck, they are combined with the scalar features, which have passed a fully connected layer before. Then, the data are passed through two decoder blocks and a final 1D-convolution.}\label{NN_figure}}
\end{figure}

We ran experiments with fully connected, convolutional, U-Net, and LSTM architectures. A slightly modified U-Net operating on atmospheric columns turned out to be the best performing type of neural network for predicting orographic gravity waves. The U-Net \citep{Ronneberger_2015}, originally developed in the context of biomedical image segmentation and widely applied for image processing tasks, consists of a contracting path (encoder) of convolutional layers, and a corresponding upsampling path (decoder), which is able to include information from the contracting path by so-called skip connections. This type of model has already proven to perform well also in the field of atmospheric science (e.g., \cite{Heuer_2024}, \cite{Gupta_2025}). Here, we adopt the principle to our problem, using one-dimensional convolutions, which are applied along the columns of winds and temperature. In the so-called bottleneck layer between contracting and up-sampling path, we inject the five scalar variables describing the orography, having passed them through one fully connected layer before. Optimising for the coefficient of determination, $R^2$, a setup with three convolutional blocks (including the bottleneck) and respective up-sampling blocks, with a total number of 2.6 million trainable parameters proves best for all our experiments. Figure \ref{NN_figure} is an illustration of this architecture.

We implement the networks in Python using the ML framework PyTorch and train for 40 epochs with a batch size of 1,024, using the \emph{Adam} optimizer, mean squared error (MSE) as loss function, a learning rate of $10^{-4}$, and \emph{LeakyReLU} as activation function. For regularisation, we apply a weight decay parameter of $10^{-5}$. 

We also tried adding an attention mechanism \citep{Vaswani_2017} to the U-Net, which did not lead to significant improvements. In addition to adding the weight decay parameter, we tried to reduce overfitting by varying the complexity of the NNs, early stopping, and by means of $L^1$ and $L^2$ penalties, batch normalisation, and drop-out layers. None of these efforts lead to significant improvements of the observed overfitting (see Section \ref{Results}).

\subsection{Experiments}

In total, we present four experiments: We consider (1) the full spectrum of gravity waves (IG) and (2) the subgrid-scale part (SG), using the same coarse input variables. Further, for both cases, we investigate (a) training the networks over all land, and (b) training only over mountainous terrain, precisely only using the 50\% of the columns with a standard deviation of the subgrid orography greater than the median standard deviation over land. We keep the architecture of the NNs fixed, but train four differnt models, independently, for the four cases. All NNs are applied to predict the GWMFs in columns over their respective training regions (all land or mountainous terrain) as a function of the coarse state variables.

\subsection{Interpretability of the Neural Networks}\label{sec_interpretability}

To ensure physical consistency and to strengthen the trust in ML models, it is important to understand how they work \citep{Heuer_2024}. This becomes increasingly difficult with more complex models. A common technique to analyse the predictions of a NN in relation to its inputs are SHAP values \citep{Lundberg_2017}, based on the game theoretic approach of \citet{Shapley_1953}. SHAP values quantify the contribution of an input feature to the model's predictions. In brief, a positive SHAP value assigned to a positive input signifies that this input pushes the prediction to a higher value. Similarly, a negative SHAP value associated with a positive input implies that the respective input decreases the predicted value, and the reverse for negative inputs. The absolute value of a SHAP value is thus a measure of the impact of a feature. If it is high, the respective variable influences the model's prediction heavily, if it is close to zero, its contribution is of minor relevance.

\section{Results}\label{Results}

\subsection{General Performance}

\begin{figure}[t]
\includegraphics[width=1.0\textwidth]{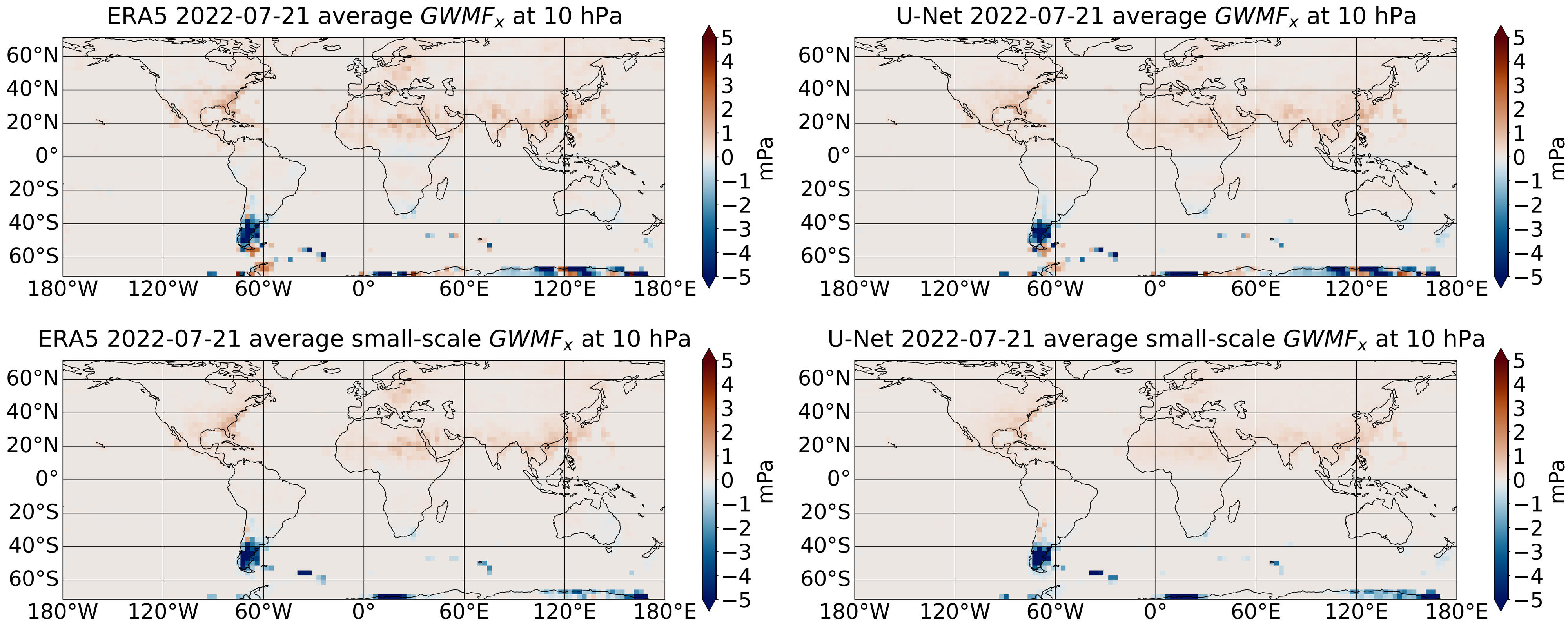}
{\caption{GWMFs in ERA5 (ground truth, left) and predictions of the U-Nets trained and applied over all land (right), for the full spectrum of gravity waves (top) and the small-scale part (bottom). The maps show fluxes averaged over 21 July 2022 at 10 hPa. Points outside the training/test regions are set to zero.}
\label{NN_predictions_land}}  
\end{figure}

\begin{figure}[t]%
\includegraphics[width=1.0\textwidth]{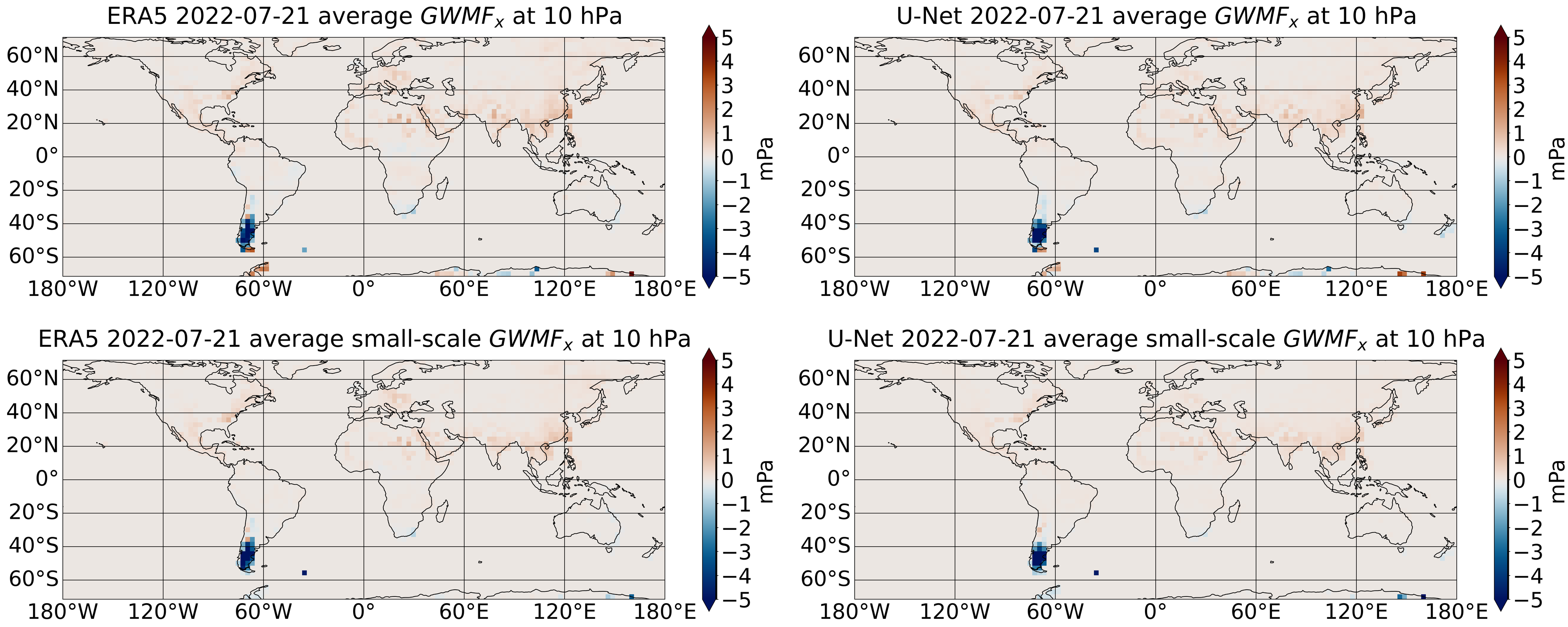}
{\caption{Like Figure \ref{NN_predictions_land}, but for the U-Nets trained and applied over mountainous terrain only.}
\label{NN_predictions_orog}}
\end{figure}
 
Figures \ref{NN_predictions_land} and \ref{NN_predictions_orog}  show GWMFs in ERA5 and predictions of the U-Nets trained and applied over all land, and only over mountainous terrain, respectively, at 10\,hPa, both for IG and SG, averaged over 21 July 2022 (24 time steps). The U-Nets are able to predict momentum fluxes of gravity waves very well, both for the full spectrum of gravity waves and for the small-scale waves. Comparing ground truth and the predictions of the NNs, we find a good agreement in structure and magnitudes globally. In particular, the NNs capture the regions of strong negative momentum fluxes, such as over the southern tip of South America around 50°S, 70°W, the belt of positive fluxes in the subtropics at 20°N of the northern hemisphere, and the positive and negative fluxes over islands in the southern Atlantic (60°S, 30°W) and the Indian Ocean (45°S, 40-70°E).

 Table \ref{experiment_table1} quantifies the skill of the U-Net in each of the four experiments in terms of the coefficient of determination, $R^2$. In general, $R^2$ values of the momentum fluxes of the full spectrum of gravity waves are higher compared to the subgrid-scale cases. We attribute this mainly to the fact that some of these waves are partially resolved on the coarse grid, such that there is information on these waves in the input data. Also, the GWMFs of the full spectrum have larger values and more pronounced structures, which seem to be easier to capture by the ML models.

Training and testing on regions with strong orography only (standard deviation greater than median) improves $R^2$ values for both IG and SG experiments, but much more for the SG setup. Although this restriction reduces the size of the training set by a factor of 2, the remaining columns seem to be more consistent in terms of the physical mechanisms related to orographic gravity waves, which the NNs are supposed to describe. In particular, we presume that the data contain less contamination from gravity waves of non-orographic sources, and potentially less noise from other atmospheric processes, leading to the observed improvements.

\begin{table}[b]
\caption{Overview of the skill of various setups of NNs on training set and test set. For every experiment, we state the values of the coefficient of determination, $R^2$, calculated over the respective training regions. We applied the NNs either over all land or only over mountainous terrain (standard deviation of the subgrid orography greater than its median over land), either for IG, or only for SG.} \label{experiment_table1}
\centering

\begin{tabular}{l c c c c}

& \textbf{$R^2$ training set} & \textbf{$R^2$ test set}  \\

\hline
IG all land model & 0.82 & 0.69 \\
IG mountains only model  & 0.86 & 0.72 \\
SG all land model & 0.70 & 0.56 \\
SG mountains only model & 0.78 & 0.63 \\
\hline
\end{tabular}
\label{tab1}

\vspace{0.2in}

\caption{Skill of the four NNs applied over different test regions. For every experiment, we state the $R^2$ values calculated on the test sets of all land, mountains only (standard deviation of the subgrid orography greater than its median over land), and flat land (standard deviation of the subgrid orography lower than its median over land).} \label{experiment_table2}
\centering
\begin{tabular}{l c c c}

\textbf{Test over ...}& \textbf{all land} & \textbf{mountains} & \textbf{flat land} \\

\hline
IG all land model & 0.69 & 0.69 & 0.60 \\
IG mountains only model & 0.62 & 0.72 & 0.43 \\
SG all land model & 0.56 & 0.58 & 0.33 \\
SG mountains only model & 0.50 & 0.63 & 0.21 \\
\hline
\end{tabular}
\label{tab2}
\end{table}

This intuition can be verified by considering the skill of the schemes trained over all land, but scored only over regions of strong orography. As seen in Table \ref{experiment_table2}, for both IG and SG, the all land schemes perform almost equivalently to the schemes trained only over strong orography. The poorer overall skill is due to the difficulty of capturing gravity wave effects over flat land. This is shown in the third column of Table \ref{experiment_table2}. While we focus on the mountains only region and scheme for the remainder of the study, the results are equivalent if we use the all land scheme.

Also, we note the difference between the training and test $R^2$ values in all experiments, which could be an indication that the NNs overfit the training data. We tried to reduce this employing various common techniques, including complexity reduction, early stopping, $L^1$ and $L^2$ penalties, batch normalisation, and drop-out layers (see Section \ref{NN_architecture}), all of which had only little impact. As described, we only kept a weight decay parameter. Since reducing the complexity (i.e. the number of trainable parameters) of the NNs leads to worse results for both training and test set, we believe the poorer skill could reflect low frequency variability of gravity wave processes, or a class imbalance, or sample size effects in the test set. Substantially more data may be needed to fully close this gap between training and test set, which is not feasible in the current study due to computational constraints. This conclusion is supported by a series of tests, where we successively increased the amount of training data. With more data, the general performance improved, and the difference between the $R^2$ on training and test sets decreased monotonically. Overall, these observations underline the intricacy of the problem, arising from the complexity and the intermittency of orographic gravity waves.

\begin{figure}[t!]
\includegraphics[width=1.0\textwidth]{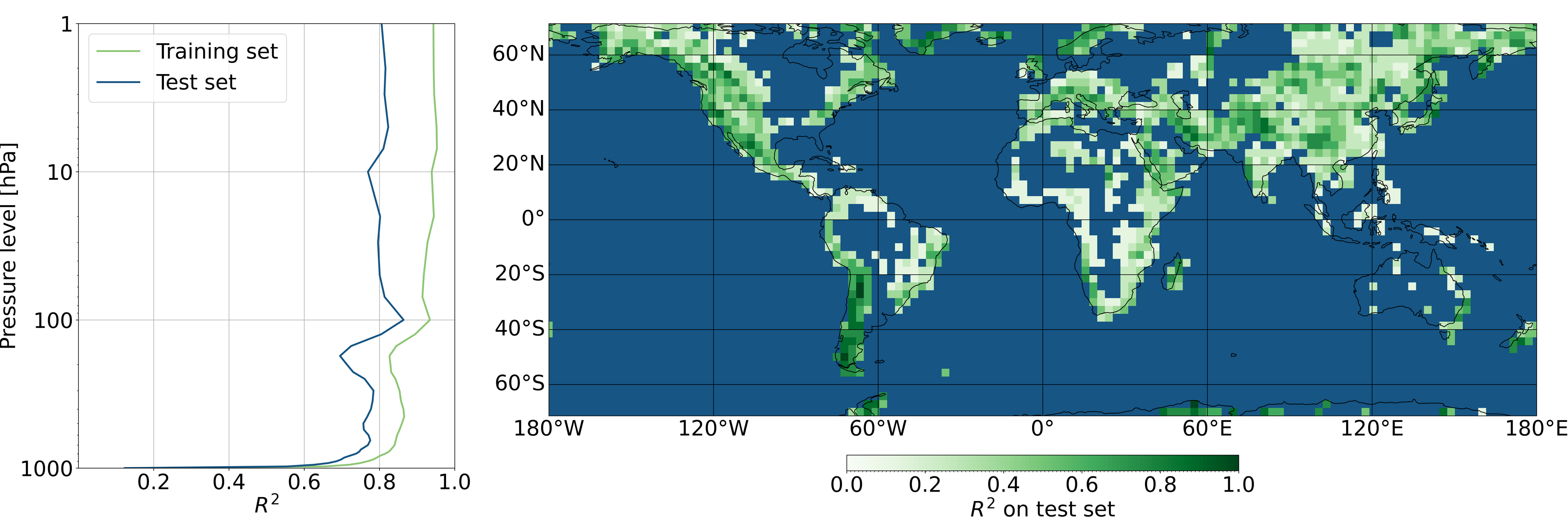}
{\caption{$R^2$ values of the U-Net trained and applied over mountainous terrain for the SG case. The left plot shows $R^2$ values of training and test set for all grid cells and time steps on various model levels, the right plot the $R^2$ values of the test set for all levels depending on the region. Grid cells outside the training/test regions are shown in dark blue; 1.5\% of the ``active" grid cells have negative $R^2$ values. }
\label{fig_levelR2}}
\end{figure}

Figure \ref{fig_levelR2} shows the performance of the NNs as a function of atmospheric pressure (left) as well as in different regions (right) for the SG case, trained and applied over mountainous orography. We find that the NNs perform similarly well at all levels, except for a minor drop of $R^2$ around 200\,hPa. This is near the tropopause, and we expect the fluxes to change significantly at this level. We also observe decreasing performance at lower heights below 500\,hPa. Generally, the quality of the predictions is better and more stable in the stratosphere (above $\sim$ 100\,hPa). This could be related to the fact that in the troposphere, convective fluxes could contaminate the data, compared to a higher importance of the local wind related to wave breaking in the stratosphere. 

Considering the horizontal distribution of $R^2$ values, the quality of the predictions is spread quite heterogeneously over the globe. The skill of the NNs is best over mountainous regions, such as the Andes, the Rocky Mountains, Greenland, and Scandinavia. For the IG case, the NNs show quite similar behaviour in terms of spatial distribution, but perform better on average (Appendix, Figure \ref{fig_levelR2_land}, see also \ref{experiment_table1}). Also, for the full spectrum, we observe a smaller decrease in performance in the lower atmosphere, where the $R^2$ of the test set basically stays around 0.8 all the way down to the ground.

\subsection{SHAP Analysis}\label{section_SHAP}

\begin{figure}[t]
\includegraphics[width=1.0\textwidth]{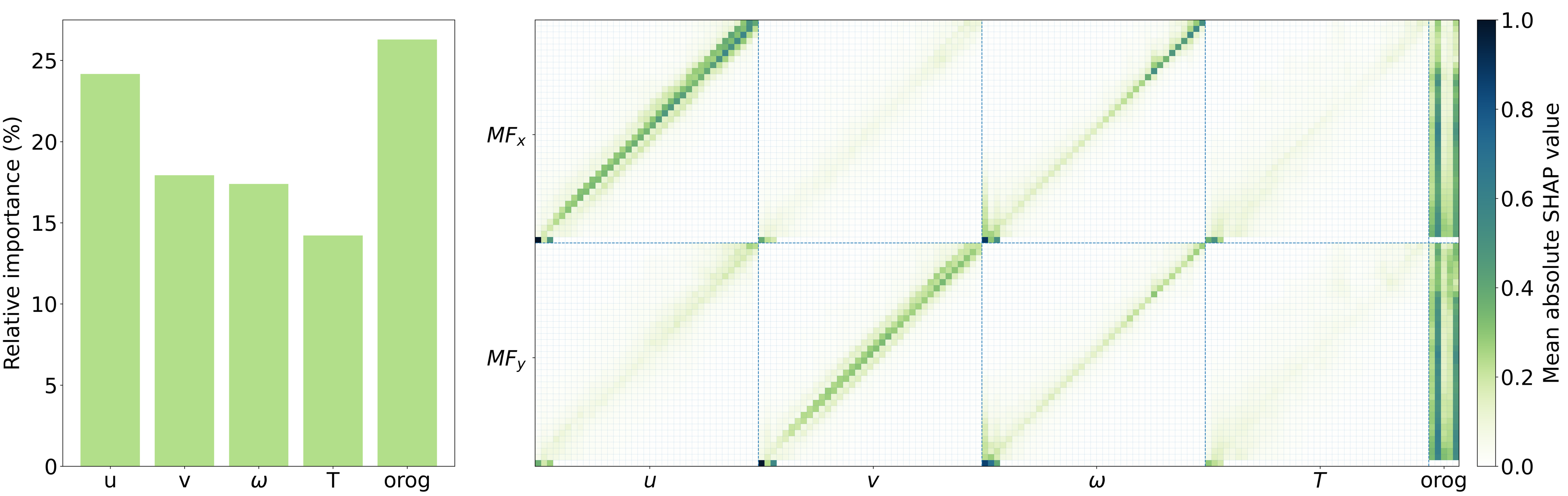}
{\caption{Relative importance of variables $u$, $v$, $\omega$, $T$, and orographic variables $z$, $\mu$, $\gamma$, $\sigma$, $\theta$, summed over all levels (left) and mean absolute SHAP values for all levels separately (right), for the SG case with the U-Net trained over mountainous terrain. In the right plot, each square depicts the relation of one of the two target variable classes $\mathrm{MF}_x, \mathrm{MF}_y$ and one of the four feature variable classes $u$, $v$, $\omega$, $T$, for all combinations of model levels; in each square, the height $z$ of the respective model level is decreasing from top to bottom and increasing from left to right. The boxes on the very right (\emph{orog}) show SHAP values of the orographic variables $z$, $\mu$, $\gamma$, $\sigma$, $\theta$ (columns from left to right). All values are normalized to $1$ by dividing by the maximal value.}
\label{shapSG}}  
\end{figure}

\begin{figure}[t]
\includegraphics[width=1.0\textwidth]{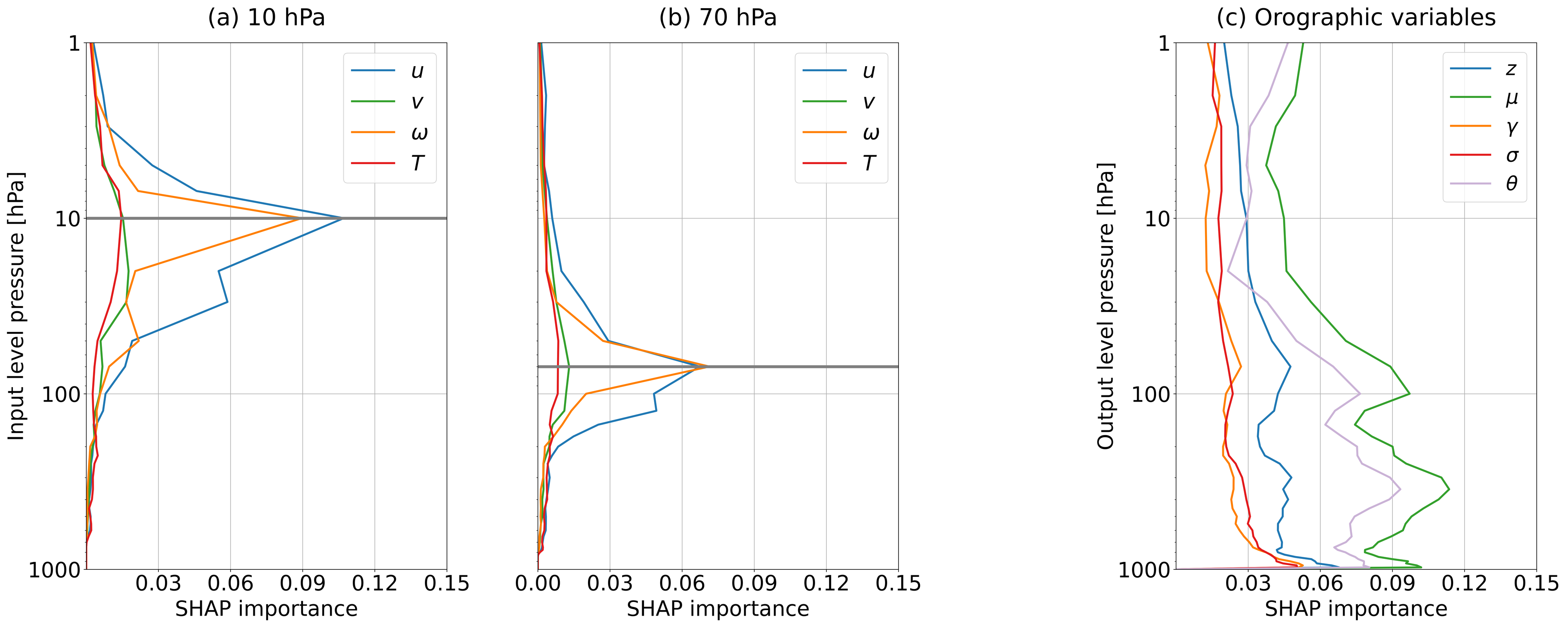}
{\caption{Absolute SHAP values for the prediction of zonal GWMFs in the SG case with the U-Net trained over mountainous terrain, averaged over all locations and times. The left plots show the SHAP importance of the input variables $u$, $v$, $T$, and $\omega$ at different pressures (y-axis) for the prediction of zonal GWMFs at (a) 10\,hPa and (b) 70\,hPa (indicated by the horizontal lines). (c) shows the SHAP importance of the orographic variables $z$, $\mu$, $\gamma$, $\sigma$, and $\theta$ for the prediction of zonal GWMFs at different pressures (y-axis).}
\label{shapSG_lev}}  
\end{figure}

A key question for the trust in our NNs is whether the predictions are physically meaningful. To test this, we analyse the networks using SHAP values (see Section \ref{sec_interpretability}). Figure \ref{shapSG} shows SHAP values for the U-Net trained with GWMFs of small-scale waves over all land, but the results are similar for all experiments.

First, we observe that the orographic variables and zonal winds $u$ constitute the most important feature classes for prediction of GWMFs, followed by meridional and vertical winds, $v$ and $\omega$, and temperature $T$. Note that in the bar plot of Figure \ref{shapSG}, $u$, $v$, $\omega$, and $T$ relate to full columns of the respective feature (37 variables each), while for orographic variables, \emph{orog} relates to five variables only. Thus, orographic features, especially standard deviation $\mu$ and slope $\theta$ are most important when comparing single inputs. While in the SG case considered here, the class of orographic features is slightly more important than the zonal winds $u$, the winds contribute strongest in the IG case, with percentages still being quite similar (Appendix Figure \ref{shapIG}).

Taking a closer look at the relations between feature and target variables, the pronounced diagonals in the right panel of Figure \ref{shapSG} clearly show that, picking a specific level of zonal momentum fluxes $\mathrm{MF_x}$, $u$ and $\omega$ on this level, as well as $\mu$ and $\theta$, are features of dominant importance. In turn, for meridional momentum fluxes $\mathrm{MF_y}$, this holds for $v$ and $\omega$. In both cases, neighbouring levels of $u$ and $v$, especially the levels just below the considered level, also contribute significantly, as can be seen in the secondary diagonals, an effect intensifying in the upper levels.

This can be seen in greater detail in Figure \ref{shapSG_lev} (a) and (b), which show the importance of the variables $u$, $v$, $\omega$, and $T$ on different levels for the prediction of zonal GWMFs at (a) 10\,hPa and (b) 70\,hPa. Similar results can be found in \cite{Connelly_2024}, and are consistent with our physical understanding of the mechanisms of gravity wave breaking and the related deposition of momentum. As gravity waves are propagating primarily upward, we expect physical relationships below and at the level of momentum deposition. The background wind at each respective level plays a decisive factor for critical level filtering. The levels above a given output level are also assigned higher SHAP values, suggesting that the vertical shear plays an important role in momentum deposition. Non-zero SHAP values well above the prediction level could be related to secondary wave generation, but could also indicate that the U-Net has learned to use correlated, but not causal relationships. The SHAP values of the five scalar orographic variables are shown in the the rectangles on the very right of Figure \ref{shapSG} and, in more detail, in Figure \ref{shapSG_lev} (c). Especially standard deviation $\sigma$ and slope $\theta$ show significant contributions throughout the atmosphere. However, we found that the correlation between $\sigma$ and $\theta$ is 0.98. The SHAP values of all features are weakening in higher regions. The corresponding plots for merdional GWMFs, and the IG case can be found in the Appendix (Figures \ref{shapSG_lev_m}-\ref{shapIG_lev_m}).

\subsection{Comparison with Lott \& Miller 1997 Parameterisation Scheme}

The main application of the NNs developed in this study is the parameterisation of subgrid-scale gravity waves in climate models. For this purpose, the relevant physical quantity is the gravity wave drag (GWD), i.e., the force per mass that gravity waves exert on the atmosphere. The GWD is the pressure gradient of the momentum flux, multiplied by the gravitational acceleration $g$:
\begin{align*}
    \mathrm{GWD}_x &= g \frac{\partial \mathrm{MF}_x}{\partial p} \\
    \mathrm{GWD}_y &= g \frac{\partial \mathrm{MF}_y}{\partial p}
\end{align*}
We investigate how our scheme compares to the GWD parameterisation developed by Lott and Miller \citep{Lott_1997, Lott_1999}, which is adopted in many models, including the ICON model \citep{Zängl_2015, Giorgetta_2018}. For our purposes, we recreated the FORTRAN implementation in Python, introducing slight simplifications and building an interface enabling the processing of ERA5 data. A more detailed description of the modifications can be found in Appendix \ref{Appendix_Lott}. The original scheme is designed to parameterise tendencies due to three processes associated with unresolved topography: the gravity wave stress as a function of height over the full atmospheric column, low-level drag associated with blocked flow, and mountain lift. As we focus only on the tendencies due to GWD, we ignore the other two outputs of the scheme for the purposes of this comparison, setting the respective tuning parameters to zero. However, even when switching these components on, they had almost no influence. Since the original scheme is tuned for the use in ICON and not for ERA5, we also tuned the parameterisation to get the best, i.e. most similar to ERA5, results achievable in terms of GWD. For details, see Appendix \ref{Appendix_Lott}.

\begin{figure}[t!]
\includegraphics[width=1.0\textwidth]{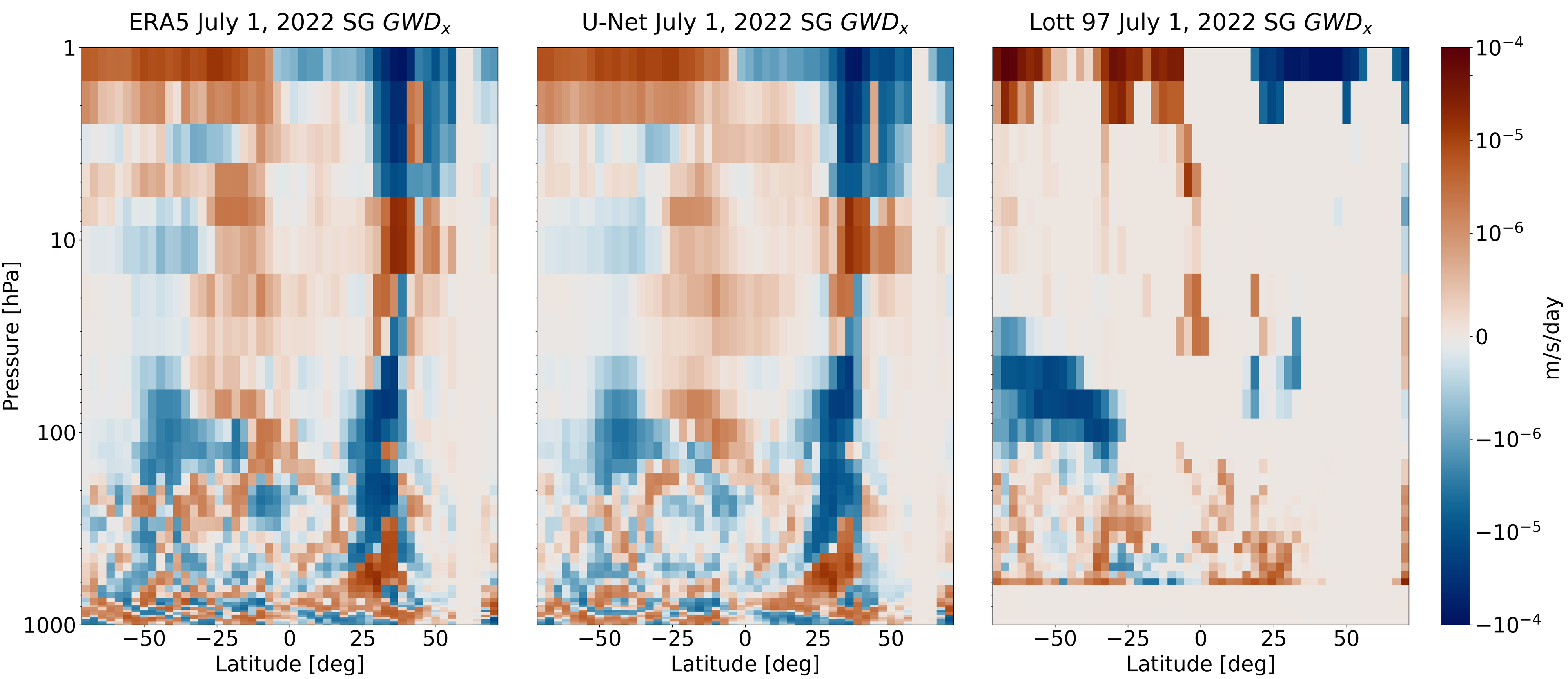}
{\caption{Zonal means of zonal gravity wave drag ($\mathrm{GWD}_x$) due to small-scale (subgrid) gravity waves in ERA5 (left), the prediction of the U-Net (middle) and the modified conventional Lott and Miller scheme (right). Data are averaged over 1 July 2022 (24 time steps).}
\label{Lott_comp}}  
\end{figure}

Figure \ref{Lott_comp} shows zonal means of zonal GWD due to small-scale waves averaged over 1 July 2022, for the ground truth ERA5 data, the prediction of our U-Net, trained on the SG case over mountains, and the conventional parameterisation. In accordance with the results above, the U-Net (middle) captures the structures of the ERA5 data (left) very well, also matching in terms of magnitudes, with only some slight local differences (e.g., missing negative fluxes at $\sim2\,\mathrm{hPa}$, $10^\circ$, and at $\sim100\,\mathrm{hPa}$, $10^\circ$, missing positive fluxes at $\sim2\,\mathrm{hPa}$, $30^\circ$, as well as invented negative fluxes at $\sim10\,\mathrm{hPa}$, $10^\circ$).

Many of the large-scale structures visible in ERA5 also appear in the conventional parameterisation's output (right). First, there is strong positive GWD in the uppermost levels of the atmosphere between the south pole and the equator in ERA5 and the Lott and Miller parameterisation. Second, we observe strong negative GWD in the NN predictions as well as in Lott and Miller's scheme at the highest levels between the equator and the north pole. as well as an attached, slightly tilted band of negative fluxes ranging from the top of the atmosphere down to around 100\,hPa. This band is interrupted by a region of positive drag (ERA5 and U-Net at $10\,\mathrm{hPa}$), while the drag simply vanishes in the conventional scheme. 

In the lower atmosphere, below 100\,hPa, there are more heterogeneous structures with positive and negative areas of GWD in both schemes. Lott and Miller's parameterisation captures some of the bigger patterns, such as the negative fluxes at around 100\,hPa poleward from $-25^\circ$ latitude and positive fluxes at $\sim75^\circ$ latitude above the ground, but it is not able to reproduce the details found in ERA5.

In summary, the predictions of our U-Net are in strong agreement with ERA5 and match it clearly better than the conventional parameterisation, which is also reflected in the respective $R^2$ values (0.76 for the U-Net, -0.36 for the conventional scheme). This is somewhat to be expected, as the U-Net was trained to capture all of ERA5's ``subgrid" waves, while Lott and Miller target only orographic waves, and also finer scales not resolved in ERA5. The poor $R^2$ of -0.36 for the conventional scheme reflects the lack of smaller scale detail. The scheme was designed to capture the gross features in the distribution of the GWD, which are more similar. This is encouraging, even if we point out that the comparison with the conventional scheme is made on a qualitative basis. A more thorough quantitative analysis will be possible as soon as the parameterisation is coupled to a climate model.

In terms of runtime, the U-Net and our Python version of the conventional parameterisation are comparable. Estimating GWD for an entire day (24 time steps) globally in the example case shown here took 103.6 seconds for the U-Net and 129.1 seconds for the conventional parameterisation. Test runs of other days lead to similar results.

\section{Discussion}

Before concluding, we summarise some of the strengths and limitations of this study and comment on how our approach compares to the study by \cite{Gupta_2025}.

First, using ERA5, we have a large and well-known dataset, with one-hourly global output over one year. However, at the resolution of $0.25^\circ$, corresponding to $\sim30\,\mathrm{km}$ at the equator, a large part of the gravity wave spectrum, i.e., waves with wavelengths smaller than $\sim200\,\mathrm{km}$, is still not resolved. In particular, many waves which need to be parameterised in climate models operating at resolutions below $100\,\mathrm{km}$ are not explicitly resolved. Therefore, our study shows that in principle, gravity wave effects can be estimated by a data-driven model, but this approach still has to be extended for data sets based on higher-resolution integrations. Also, the highest level in our setup is at 1\,hPa, which corresponds to $\sim50\,\mathrm{km}$, while state-of-the art climate models usually work with a higher model top, and gravity waves particularly impact the dynamics of the stratosphere and mesosphere. 

Second, while the filtering with MODES is the physically most precise approach, since it respects the dynamical properties of gravity waves, the software does not differentiate GWs with respect to their sources. This work focuses on orographic gravity waves, while being aware that despite considering only regions over land, respectively over mountains, there is contamination by the effects of non-orographic waves. Also, using full $\omega$ for calculation of the fluxes possibly incurs some noise from other sources. 

Finally, we want to point out the differences of our approach and the work of \cite{Gupta_2025}, which also employs ERA5 data to train NNs for GWMFs. Opposed to our approach, the authors aim at including horizontal propagation of gravity waves, using a U-Net with attention mechanism operating on the full 3D model state and predicting full 3D gravity wave fields. While this eliminates one of the severest shortcomings of gravity wave parameterisations, coupling such a non-local scheme to a climate model is challanging. The column approach taken in our work makes it possible to use significantly smaller networks (2.6 million trainable parameters, compared to 38 million in \cite{Gupta_2025}), and would allow integration in current climate model parameterisation frameworks. In addition, using winds, temperature, and topographic parameters, we are able to give a physical interpretation of the predictions of our NNs in terms of SHAP values, as shown in Section \ref{section_SHAP}.

\section{Conclusions}

This study demonstrates that NNs can predict the force exerted by gravity waves over topographic regions. Employing the full year 2024 of the ERA5 reanalysis, momentum fluxes of gravity waves were calculated in the original ERA5 resolution, and then coarsened for the training of various ML models designed to predict the flux based on the coarsened model state alone. Four different NNs estimate momentum fluxes for either the full or the subgrid-scale gravity wave spectrum, over all land or over mountainous terrain only. A U-Net acting on atmospheric columns, with scalar features injected in the bottleneck layer, proved to perform best for this task in an offline setting.

Evaluating the networks on selected dates from the year 2022, we found that the full spectrum of gravity waves was easier to predict compared to only the subgrid-scale waves ($R^2$ values 0.69 and 0.56 for the full and subgrid spectra, respectively, on the test set over all land). We attribute this difference in performance mainly to the fact that a part of the full spectrum of gravity waves is still resolved in the coarse resolution, and that the subgrid-scale part has more fine-scale structure. Training and testing on only regions with mountainous terrain, or nearly equivalently, scoring the all land models only over mountainous regions, improves the performance ($R^2$ values 0.72 and 0.63 on the test set), especially in the subgrid-scale case. These results suggest the GWMFs over flatter land are associated with other processes (e.g., convection, fronts), which are not well represented in the input featrues to our models, targeted to orographic gravity waves. An issue still to be addressed is that our networks appear to overfit the training data. This might be improved by use of a larger and more diverse dataset, and will be investigated in future work.

Analysis of the NNs with XAI methods reveals that the contribution of input features is in accordance with our understanding of the physical mechanisms of orographic gravity waves. In particular, variables describing the orography as well as zonal (meridional) wind, and vertical wind are the dominant features for predicting zonal (meridional) momentum fluxes. Considering GWMFs on a given level, winds at the same and at neighbouring levels are most important, consistent with the role of dissipation by critical layers in our theoretical understanding of gravity wave dynamics.

Finally, a comparison with the conventional, physics-based parameterisation by \cite{Lott_1997} and  \cite{Lott_1999} in terms of the gravity wave drag shows a good qualitative agreement. Key structures of the drag are captured in both approaches, which further supports the credibility and physical consistency of the NNs. However, the NNs capture the variability of the high-resolution data substantially better.

In a next step, we will extend our NNs to the case of non-orographic gravity waves. By including additional feature variables related to sources like convection, jets, or fronts, and training also over oceans, we aim for a scheme addressing the various wave types all in one. Also, we are investigating techniques of transfer learning \citep{Gholizade_2025} to adapt our parameterisation to the ICON XPP model \citep{Mueller_2025} and run coupled online simulations. A question still to be answered is how to tackle horizontal propagation of gravity waves, which is not covered by the current one-column approaches and poses challenges in coupling the ML-based parameterisation to the climate model. Nonetheless, this work demonstrates the potential of data-driven approaches to significantly improve the representation of gravity waves in models.

This study shows the capabilities of ML to accurately estimate gravity wave effects in weather and climate models and thereby improve their predictions and projections. Even as the resolution of models increases in the future, parameterisations will still be relevant for running large ensembles. To this end, for physically complex, non-local, transient, and multi-scale phenomena like gravity waves, NNs can play a key role. In addition, precise representations will help to improve our physical understanding, especially regarding the effects of gravity waves in the complex Earth system and their behaviour in a changing climate.

\vspace{0.4in}

\textbf{Acknowledgments}\quad  The authors thank the entire matrix group \emph{Middle Atmosphere} at the Institute of Atmospheric Physics, Deutsches Zentrum für Luft- und Raumfahrt e.V., Oberpfaffenhofen, for fruitful discussions, constructive ideas, and their support. Also, we thank Claudia C. Stephan, Leibniz Institute of Atmospheric Physics at the University of Rostock, Ostseebad Kühlungsborn, for interesting conversations, her input and assistance.

\textbf{Funding}\quad  Funding for this study was provided by the European Research Council (ERC) Synergy Grant “Understanding and Modelling the Earth System with Machine Learning (USMILE)” under the Horizon 2020 research and innovation programme (Grant agreement No. 855187). The authors gratefully acknowledge the Earth System Modelling Project (ESM) for funding this work by providing computing time on the ESM partition of the supercomputer JUWELS (Jülich Supercomputing Centre, 2021) at the Jülich Supercomputing Centre (JSC). M.S. acknowledges support from the DLR Quantum Computing Initiative and the Federal Ministry of Research, Technology and Space. E.G. acknowledges support from the US NSF through award OAC-2004572, and Schmidt Sciences, as part of the Virtual Earth System Research Institute (VESRI). V.E. was additionally supported by the Deutsche Forschungsgemeinschaft (German Research Foundation) through the Gottfried Wilhelm Leibniz Prize awarded to V.E. (Reference No. EY 22/2-1).

\textbf{Author contributions}\quad Conceptualisation: E.H.; M.S.; A.D.; M.R.; V.E. Methodology: E.H.; M.S.; A.D.; E.G.; M.R.; N.Z.; V.E. Data curation: E.H. Data visualisation: E.H. Writing original draft: E.H.; M.S.; V.E. All authors approved the final submitted draft.

\textbf{Data availability}\quad The code of this work will be published under \url{https://github.com/EyringMLClimateGroup/haslauer26MLE_NNs_GravityWaves_ERA5/}.
ERA5 reanalysis data \citep{Hersbach_2020} is available at \url{https://cds.climate.copernicus.eu/}. For our study, we used the datasets \emph{ERA5 hourly data on pressure levels from 1940 to present} and \emph{ERA5 hourly data on single levels form 1940 to present}. The software MODES \citep{Zagar_2015} can be  obtained upon request at \url{https://modes.cen.uni-hamburg.de/}.


\bibliography{references.bib}

\normalsize

\newpage
\begin{appendix}
\section{Appendix}\label{appendix}
\subsection{Settings for MODES}\label{MODES_settings}
For applying the software MODES \citep{Zagar_2015}, the following parameters were used:

First, describing up to which vertical and horizontal mode normal mode function decomposition is performed, we chose:

\begin{itemize}
\item Number of vertical modes: \texttt{num\_vmode\,=\,30}
\item Zonal wavenumbers: \texttt{num\_zw\,=\,400}
\item Number of Hough modes for \texttt{EIG}, \texttt{WIG}, and \texttt{ROT}: \texttt{maxl\,=\,230}
\end{itemize}

Second, for the projection back to physical space, we kept all vertical modes, all \texttt{EIG} and \texttt{WIG} modes (corresponding to inertia-gravity waves) and excluded all \texttt{ROT} modes. For the IG case, we kept all \texttt{kmode} when projecting back to physical space. For the SG, we chose \texttt{kmode\_s\,=\,0} and \texttt{kmode\_e\,=\,16}.

\subsection{Python Implementation of Lott \& Miller 1997 GWD Scheme}\label{Appendix_Lott}
Our version of the parameterisation scheme by \citet{Lott_1997} and \cite{Lott_1999} is based on the FORTRAN code used in ICON \citep{Giorgetta_2018, Zängl_2015}, namely the module \texttt{mo\_ssodrag}. We translated the code into Python as accurately as possible, with some changes to render it for the ERA5 case and to tune it to get the best results possible:
\begin{itemize}
\item All loops and the corresponding commands for parallelisation over horizontal grid points were removed. Parallelisation is implemented in ICON to reduce computation time, but is not required for our tests.
\item The parts of the scheme not touched by our parameterisation, namely blocked low-level flow drag and mountain lift, were disabled. This was done by setting the tuning parameters \texttt{gkwake} and \texttt{gklift} to zero. However, switching these parts on with comparable tuning parameters led to only very slight differences.
\item The tuning parameter \texttt{gkdrag}, which controls the strength of the gravity wave drag part of the parameterisation was set to \texttt{0.0001}, the parameter \texttt{nktopg} to \texttt{36}.
\item We built an interface preparing all relevant variables from ERA5, and passing them to the parameterisation, which yields zonal and meridional drag (in the terminology of ICON, \texttt{pdu\_sso} and \texttt{pdv\_sso}).
\item In the sense of ``tuning" the parameterisation to the ERA5 data, indices corresponding to atmospheric levels in some loops were changed from \texttt{1} to \texttt{0}, such that GWD occures also in the uppermost level.
\end{itemize}

\newpage
\subsection{Additional Figures}\label{Apx_Figs}

\renewcommand\thefigure{\thesection\arabic{figure}}    \setcounter{figure}{0}

\begin{figure}[h]
\includegraphics[width=1.0\textwidth]{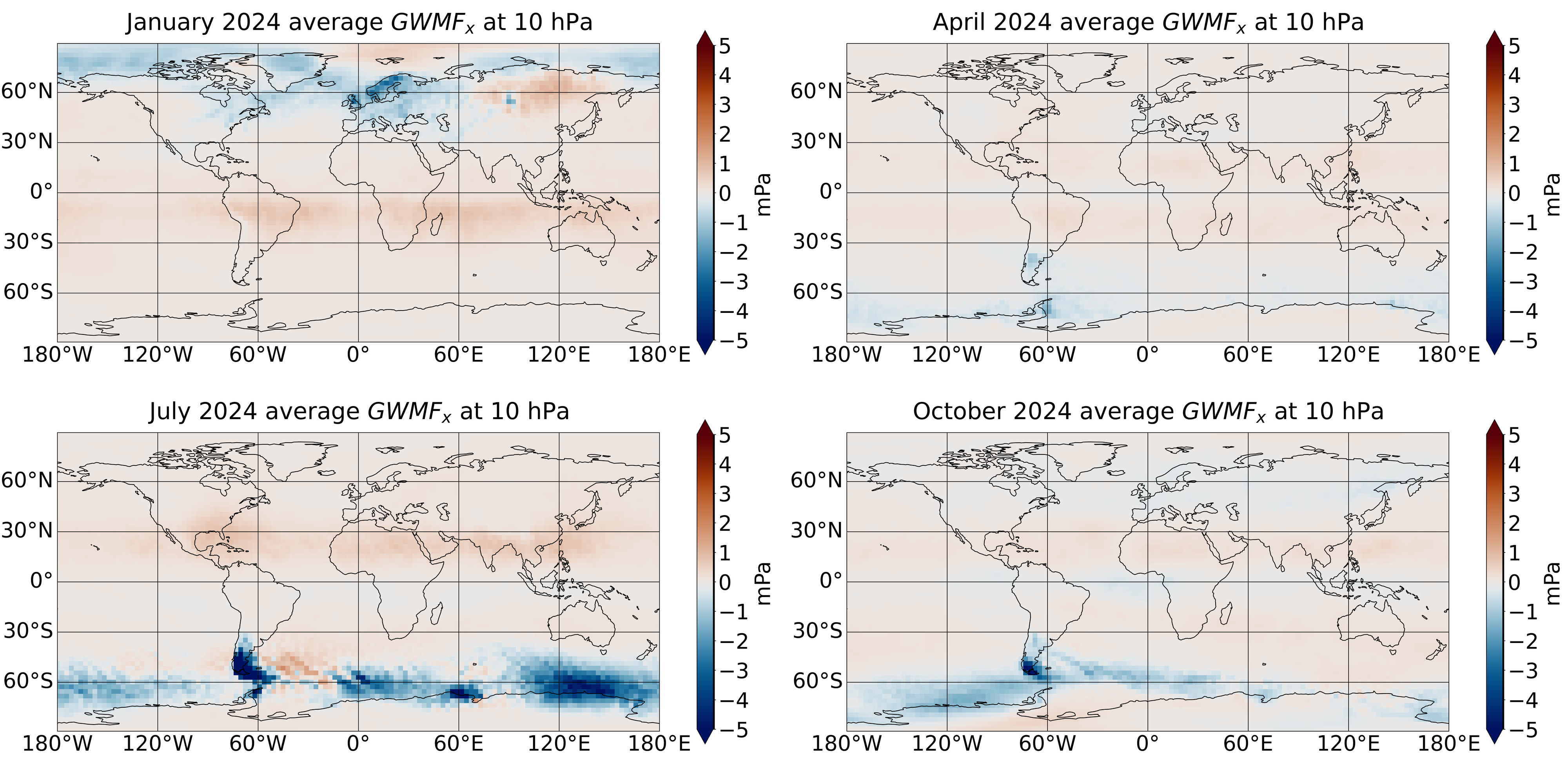}
{\caption{Monthly averages of zonal GWMFs in ERA5 of the full spectrum of gravity waves for January, April, July, and October 2024.}
\label{MF_IG_seasons}}
\end{figure}

\begin{figure}
\includegraphics[width=1.0\textwidth]{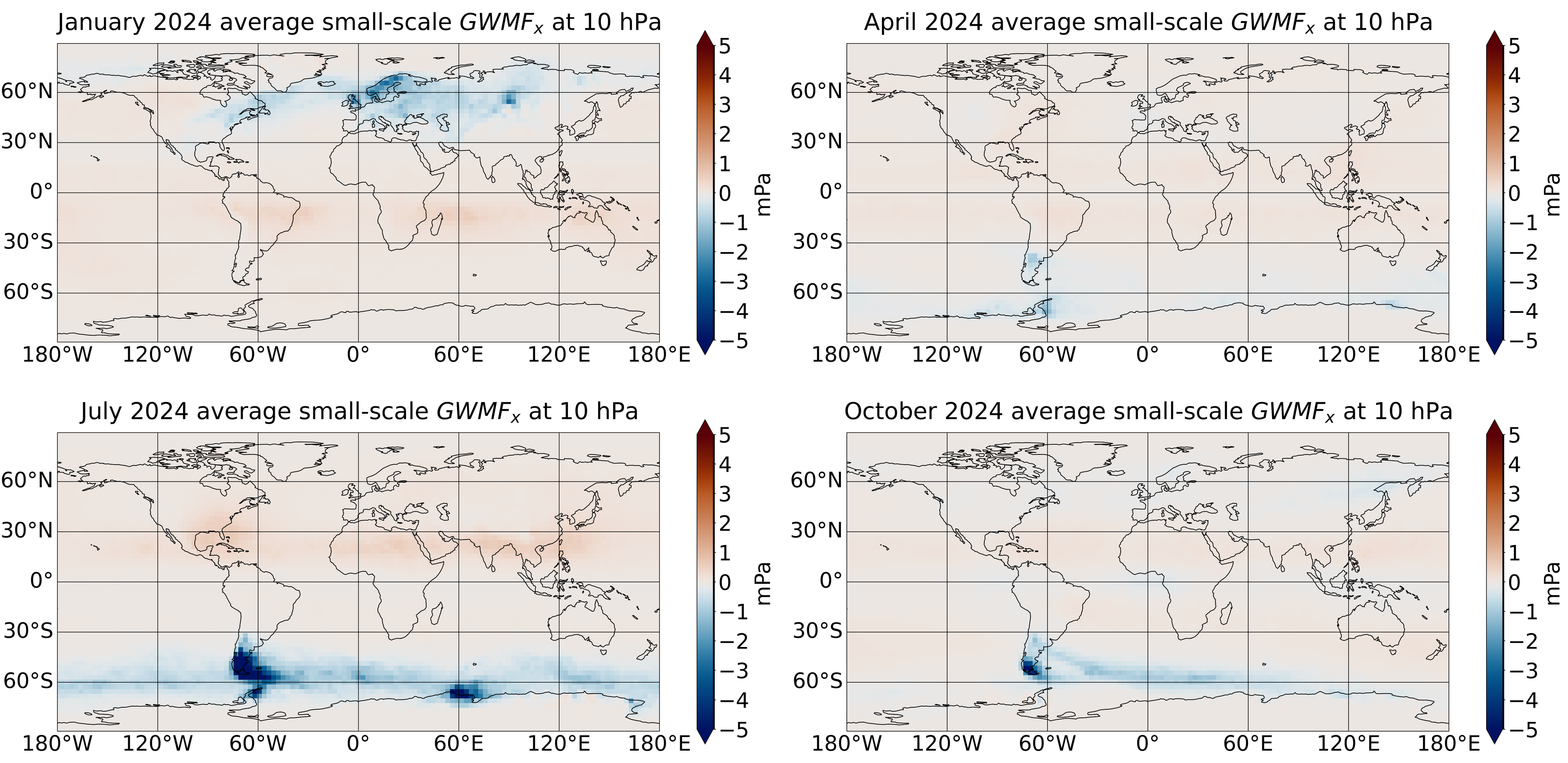}
{\caption{Like Figure \ref{MF_IG_seasons}, but for the small-scale part of the gravity wave spectrum.}
\label{MF_SG_seasons}}  
\end{figure}

\pagebreak

\begin{figure}
\includegraphics[width=1.0\textwidth]{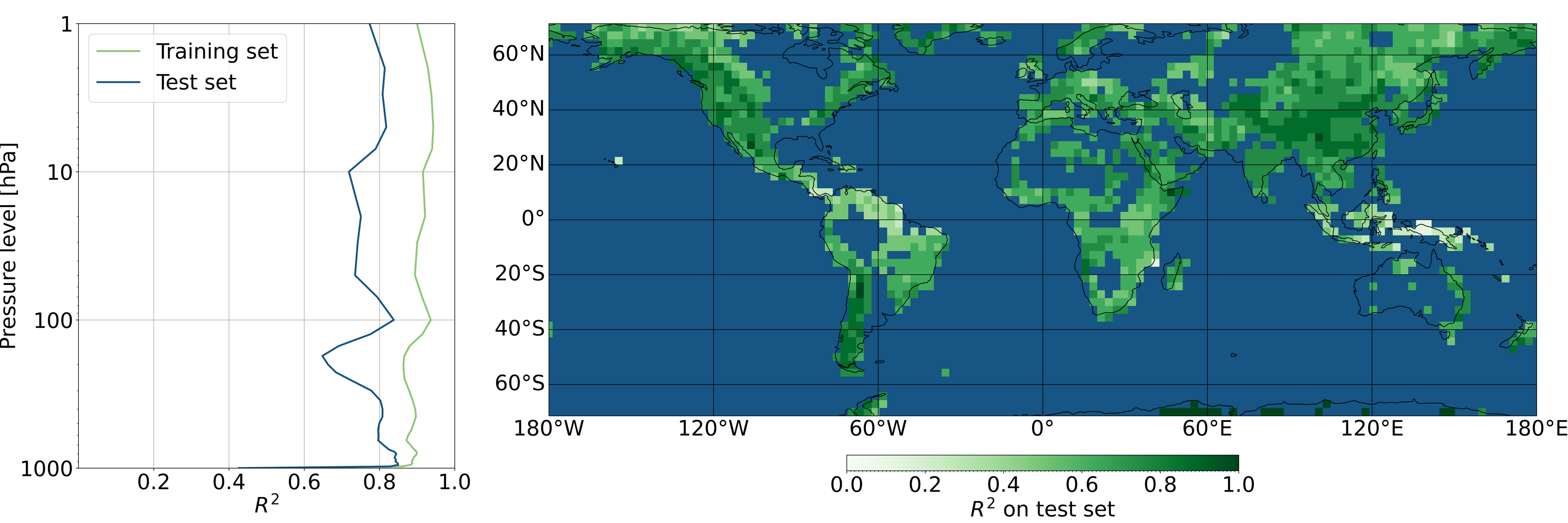}
{\caption{$R^2$ values of the U-Net trained and applied over mountainous terrain for the IG case. The left plot shows $R^2$ values of training and test set for all grid cells and time steps on various model levels, the right plot the $R^2$ values of the test set for all levels depending on the region. Grid cells outside the training/test regions are shown in dark blue; none of the ``active" grid cells have $R^2$ values below 0.}
\label{fig_levelR2_land}}
\end{figure}

\begin{figure}
\includegraphics[width=1.0\textwidth]{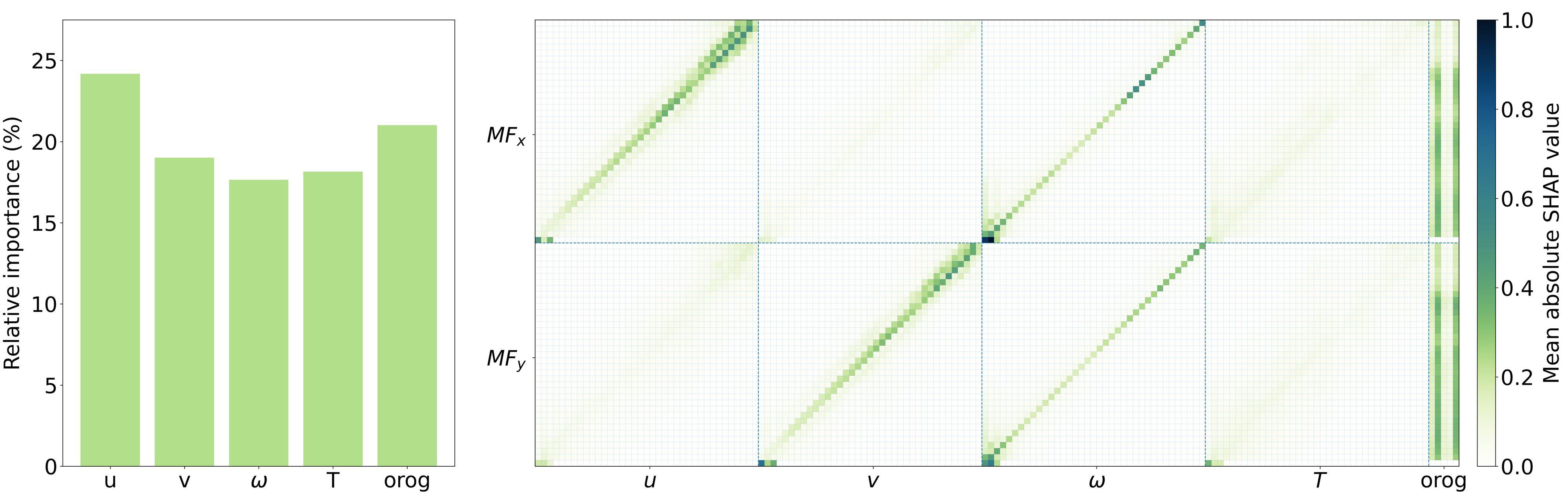}
{\caption{Relative importance of variables $u$, $v$, $\omega$, $T$, and orographic variables $z$, $\mu$, $\gamma$, $\sigma$, $\theta$, summed over all levels (left) and mean absolute SHAP values for all levels separately (right), for the IG case with the U-Net trained over mountainous terrain. In the right plot, each square depicts the relation of one of the two target variable classes $\mathrm{MF}_x, \mathrm{MF}_y$ and one of the four feature variable classes $u$, $v$, $\omega$, $T$, for all combinations of model levels; in each square, the height $z$ of the respective model level is decreasing from top to bottom and increasing from left to right. The boxes on the very right (\emph{orog}) show SHAP values of the orographic variables $z$, $\mu$, $\gamma$, $\sigma$, $\theta$ (columns from left to right). All values are normalized to $1$ by dividing by the maximal value.}
\label{shapIG}}  
\end{figure}

\begin{figure}
\includegraphics[width=1.0\textwidth]{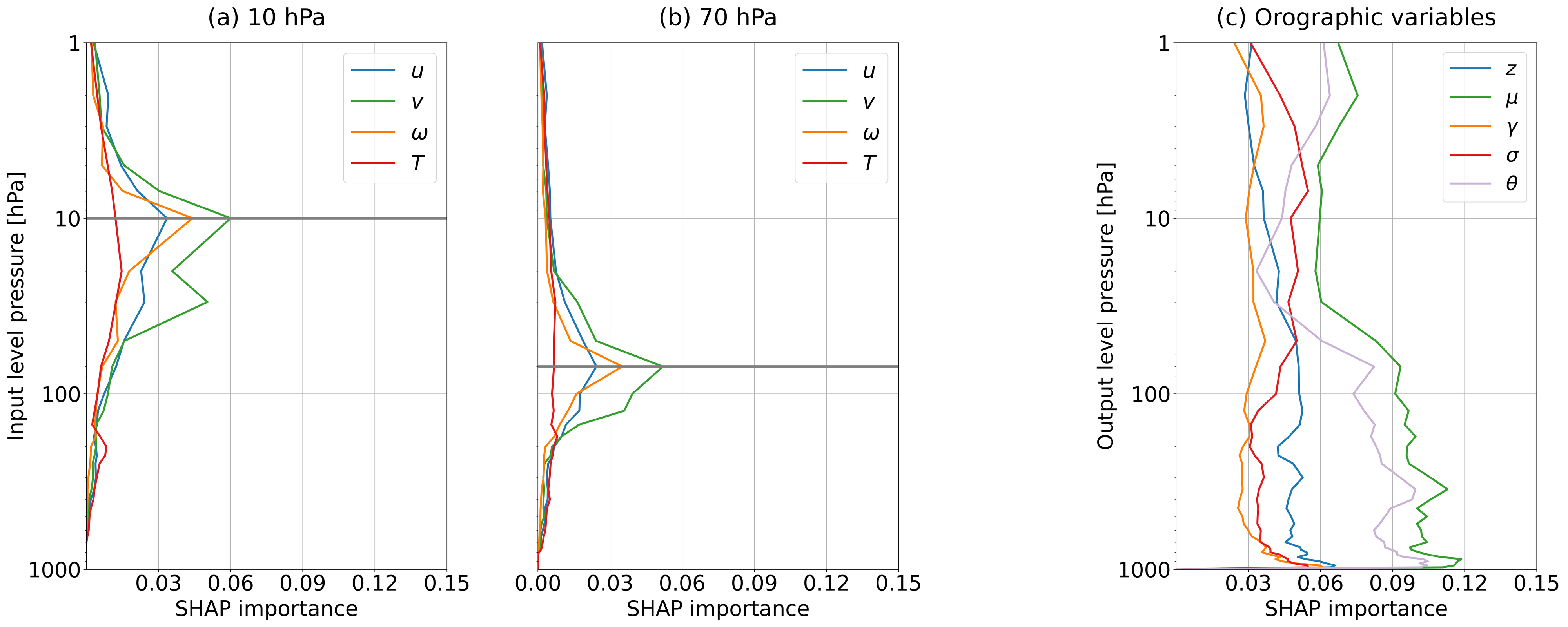}
{\caption{Absolute SHAP values for the prediction of meridional GWMFs in the SG case with the U-Net trained over mountainous terrain, averaged averaged over all locations and times. The left plots show the SHAP importance of the input variables $u$, $v$, $T$, and $\omega$ at different pressures (y-axis) for the prediction of zonal GWMFs at (a) 10\,hPa and (b) 70\,hPa (indicated by the horizontal lines). (c) shows the SHAP importance of the orographic variables $z$, $\mu$, $\gamma$, $\sigma$, and $\theta$ for the prediction of zonal GWMFs at different pressures (y-axis).}
\label{shapSG_lev_m}}  
\end{figure}

\begin{figure}
\includegraphics[width=1.0\textwidth]{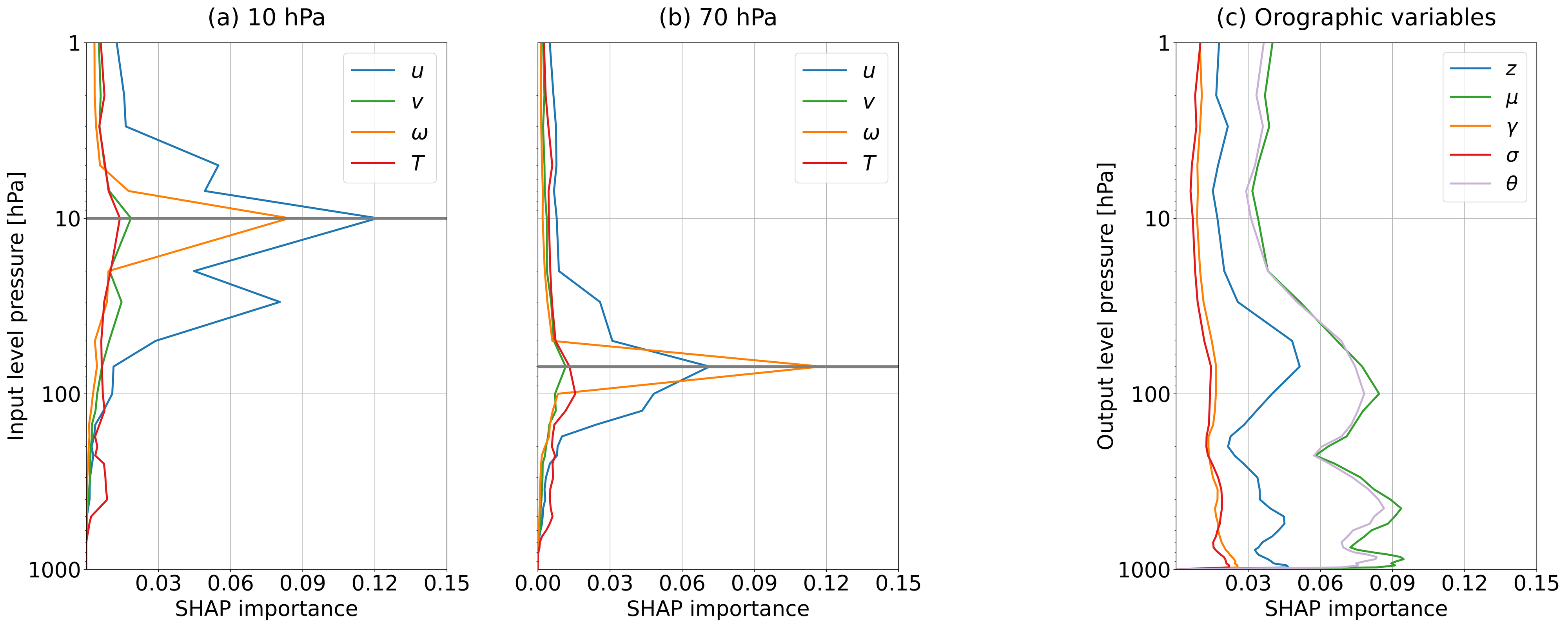}
{\caption{Like Figure \ref{shapSG_lev_m}, but for zonal GWMFs in the IG case.}
\label{shapIG_lev}}  
\end{figure}

\begin{figure}
\includegraphics[width=1.0\textwidth]{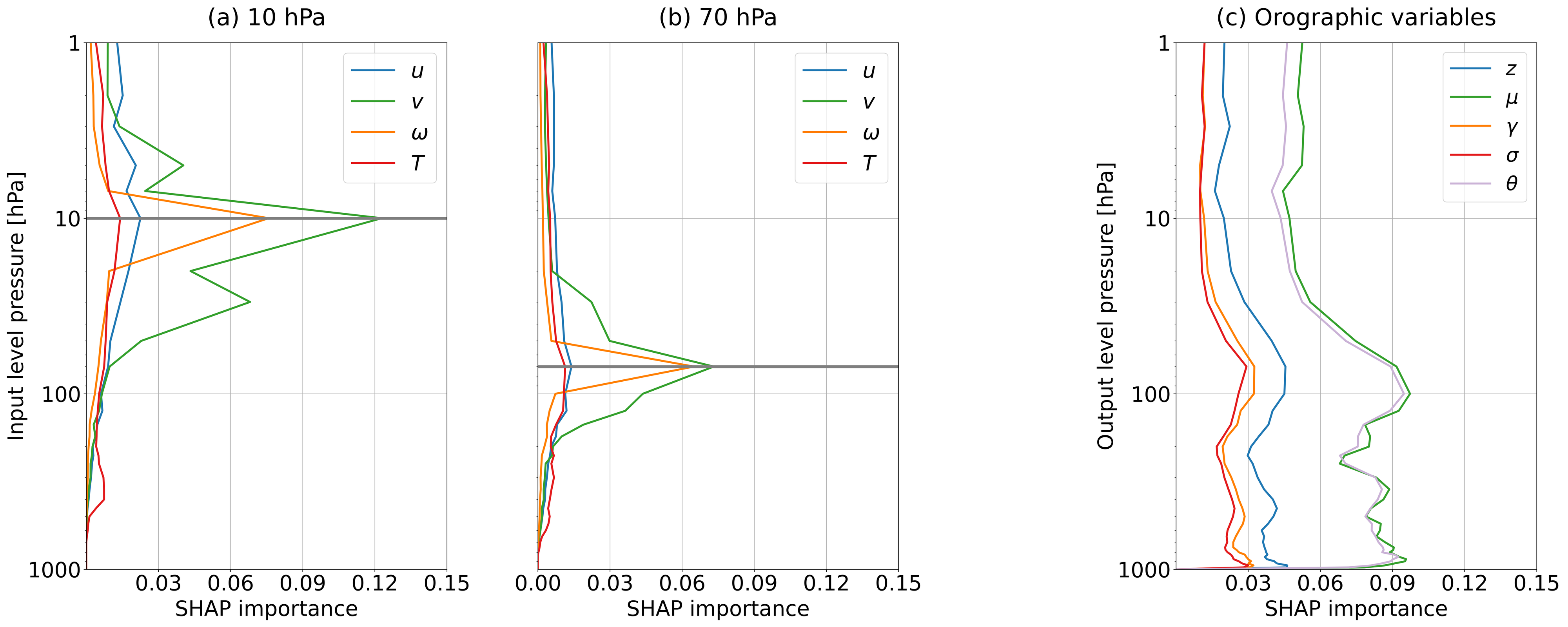}
{\caption{Like Figure \ref{shapSG_lev_m}, but for meridional GWMFs in the IG case.}
\label{shapIG_lev_m}}  
\end{figure}

\end{appendix}

\end{document}